\documentclass[aps,prl,preprint,superscriptaddress,floatfix,showpacs]{revtex4-1} 
\usepackage{graphicx} 
\usepackage[colorlinks=true,citecolor=blue,urlcolor=blue,linkcolor=blue]{hyperref} 
\usepackage{amsmath} 
\usepackage{bm} 
\usepackage{textcomp}
\begin{document}

\title{Breakdown of the Dipole Approximation in Strong-Field Ionization}

\author{A. Ludwig}
\email{aludwig@phys.ethz.ch}
\affiliation{Department of Physics, ETH Zurich, 8093 Zurich, Switzerland}
\author{J. Maurer}
\affiliation{Department of Physics, ETH Zurich, 8093 Zurich, Switzerland}
\author{B.W. Mayer}
\affiliation{Department of Physics, ETH Zurich, 8093 Zurich, Switzerland}
\author{C.R. Phillips}
\affiliation{Department of Physics, ETH Zurich, 8093 Zurich, Switzerland}
\author{L. Gallmann}
\affiliation{Department of Physics, ETH Zurich, 8093 Zurich, Switzerland}
\affiliation{Institute of Applied Physics, University of Bern, 3012 Bern, Switzerland}
\author{U. Keller}
\affiliation{Department of Physics, ETH Zurich, 8093 Zurich, Switzerland}

\date{October 2, 2014}

\begin{abstract}
We report the breakdown of the electric dipole approximation in the long-wavelength limit in strong-field ionization with linearly polarized few-cycle mid-infrared laser pulses at intensities on the order of 10$^{13}$\,W/cm$^2$. Photoelectron momentum distributions were recorded by velocity map imaging and projected onto the beam propagation axis. We observe an increasing shift of the peak of this projection opposite to the beam propagation direction with increasing laser intensities. From a comparison with semiclassical simulations, we identify the combined action of the magnetic field of the laser pulse and the Coulomb potential as origin of our observations.
\end{abstract}


\maketitle

The electric dipole approximation is a concept widely used to facilitate calculations and the understanding of processes involved in light-matter interactions in atomic, molecular and optical physics. In its essence, it assumes that the relevant length scales associated with the target are small compared to the wavelength of light. In particular, theoretical descriptions of strong-field ionization build heavily on this approximation and it usually holds well for the most commonly used near-infrared laser sources and intensities~\cite{Milosevic2006}. While the breakdown of the dipole approximation towards short wavelengths, where the wavelength becomes comparable to the target size, can be expected, a lesser-known limit also exists towards long wavelengths~\cite{Reiss2008,Reiss2013,Reiss2014}. In the dipole approximation, the vector potential $\bm{A}(t)$ that describes the laser field is spatially homogenous and thus, the magnetic field component of the laser field is zero, since $\bm{B}$=$\nabla{\times}\bm{A}(t)$=$0$. Consequently, the magnetic field component of the laser field is neglected in all considerations building on this approximation. However, because the magnetic-field component of the Lorentz force acting on the electrons exposed to the laser light depends linearly on the ratio $v/c$ with the electron's velocity $v$ and the speed of light in vacuum $c$, high-energy electrons are strongly influenced by the magnetic field. Moreover, such high-energy electrons inevitably occur in strong-field ionization using intense long-wavelength driving lasers. Thus, at long wavelength, the dipole approximation is expected to break down in strong-field ionization due to the onset of magnetic field effects.

\begin{figure}[t]
\includegraphics{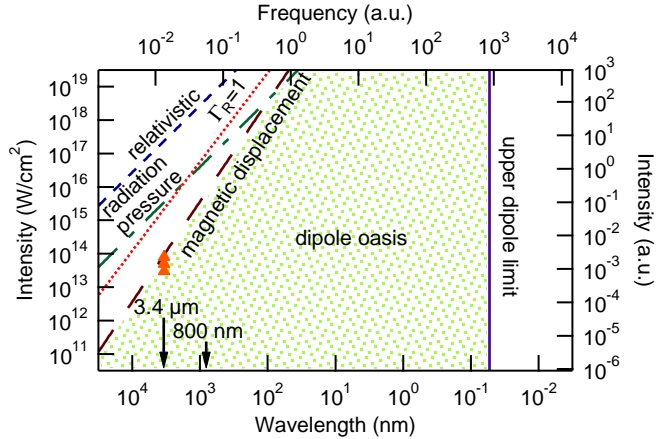}
\caption{\label{fig1}Illustration of the wavelength-intensity parameter space in strong-field ionization, taking the magnetic field component into account. The area where the dipole approximation is considered as valid (“dipole oasis”) is depicted as green dotted region. The well-known short-wavelength dipole limit arises for wavelengths on the order of the atomic scale, i.e.~for $\lambda$=1\,a.u.. The long-wavelength limit arises due to the laser magnetic field component, and is characterized by the ratio $U_\text{p}/2\omega c=$1\,a.u.. The experiment presented in this letter was performed at a wavelength of 3.4\,\textmu m at intensities close to this limit (orange triangles). The radiation pressure limit arises for $U_\text{p}^2/2c^2=$0.5\,a.u., and true relativistic effects start to occur around $2U_\text{p}/c^2=$1\,a.u..
The parameter $\Gamma_\text{R}=\sqrt{U_\text{p}^3I_\text{p}}/3c^2\omega$ indicates the limit where the spatially spread electron wavepacket essentially misses the ion under the influence of the magnetic field~\cite{Palaniyappan2006}.
Figure adapted from Ref.~\cite{Reiss2014}.}
\end{figure}

Criteria that characterize the onset of magnetic field effects as well as the onset for fully relativistic treatment of the ionization process have been formulated~\cite{Reiss2008,Reiss2013,Joachain2012,Reiss2014}. The commencement of fully relativistic behavior can be characterized by twice the ponderomotive potential approaching the rest energy of the electron, i.e., $2U_\text{p}/c^2=I/2\omega^2c^2=1$, with $I$ the peak intensity and $\omega$ the carrier frequency of the laser pulse (atomic units are used in the equations throughout this letter). The onset of the influence of the magnetic field effects, however, becomes noticeable already at significantly smaller intensities and higher frequencies than those required to achieve this condition. In particular, the limit of the dipole approximation for long wavelengths is reached when the magnetic field induced amplitude of a free electron's motion in the frame where the electron is in average at rest becomes 1\,a.u., i.e., $U_\text{p}/2\omega c=1$\,a.u.~\cite{Reiss2008,Reiss2013,Reiss2014}. These relativistic and non-relativistic limits of the dipole regime are shown in Fig.~\ref{fig1}. Due to the widespread deployment of Ti:sapphire laser systems, the majority of experiments in strong field science are performed at wavelengths around 800\,nm, where the dipole approximation is considered to be valid for intensities of at least 5$\times$10$^{15}$\,W/cm$^2$~\cite{Reiss2008}.

Non-dipole effects in strong-field ionization have been subject to a number of works by various groups. Such effects have been observed experimentally for the case of multiply charged ions in ultra-high-intensity beams at wavelengths of 800\,nm~\cite{DiChiara2010,Chowdhury2003,Palaniyappan2005} and 1053\,nm~\cite{Moore1995,Meyerhofer1996}. They have been studied purely theoretically for XUV pulses~\cite{Forre2006}, in the frame of calculations on photoelectron rescattering processes~\cite{Palaniyappan2006,Klaiber2006,Chirila2002,Walser2000}, and laser driven ion dynamics~\cite{Hu2001}.
 Furthermore, studies of non-dipole effects have often assumed a negligible influence of the Coulomb potential, as was recently the case in an experiment for circularly polarized light at 800\,nm and 1.4\,\textmu m~\cite{Smeenk2011}, and in theoretical investigations~\cite{Reiss2008,Reiss2013,Titi2012}. Here, we present an experimental study on non-dipole strong-field ionization for the important case of linearly polarized light with few-cycle pulses at a mid-infrared (mid-IR) wavelength, where the Coulomb potential of the residual ion is of significant importance as well as the laser's magnetic field. A solid understanding of the case of linearly polarized light at long wavelengths is of considerable importance for all phenomena relying on electron recollision processes such as the generation of x-ray high harmonic radiation and attosecond pulses~\cite{Popmintchev2012}, holography with photoelectrons~\cite{Huismans2011} and laser-induced diffraction~\cite{Meckel2008,Blaga2012}.

In this letter, we study non-dipole effects on complete photoelectron momentum distributions (PMDs) from strong-field ionization of noble gases with few-cycle mid-IR pulses at moderate intensities. We show that these PMDs exhibit clear evidence for the influence of the magnetic field component of the laser pulse. To access the long-wavelength limit of the dipole approximation, we developed a state-of-the-art optical parametric chirped-pulse amplifier (OPCPA) system based on chirped quasi-phase-matching (QPM) devices, as described in detail elsewhere~\cite{Mayer2013,Mayer2014}. This system delivers laser pulses with a duration of 44\,fs, and a pulse energy of 21.8\,\textmu J and a center wavelength of 3.4\,\textmu m, with a high repetition rate of 50\,kHz. The laser beam was guided into a velocity map imaging spectrometer (VMIS)~\cite{Eppink1997,Ghafur2009,Weger2013} and focused into the interaction region by a dielectric mirror with a focal length of 15\,mm.
The resulting photoelectrons were mapped onto a micro-channel plate, imaged by a successive phosphor screen and recorded with the help of a CCD camera.

\begin{figure}[b]
\includegraphics{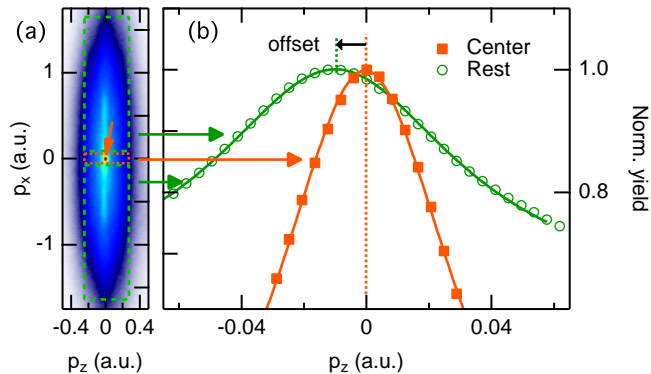}
\caption{\label{fig2}(a) Typical projected photoelectron momentum distribution (PMD) of xenon recorded at an intensity of 6$\times$10$^{13}$\,W/cm$^2$ with linear polarization using a VMIS at a center wavelength of 3.4\,\textmu m. We show the plane spanned by the laser polarization (labeled $p_\text{x}$) and propagation (labeled $p_\text{z}$) direction. The orange arrow depicts the center spot resulting from field-ionization of highly-excited Rydberg states used as reference for $p_\text{z}$=0\,a.u., and the dashed boxes indicate the areas taken for the momentum-offset analysis. (b) Projections of the PMD onto the beam propagation direction together with Lorentzian fits. The orange curve (squares) is used to set the $p_\text{z}$=0\,a.u. reference and the offset of the maximum of the photoelectron distribution is extracted from the fit on the green markers (circles).}
\end{figure}

We recorded PMDs from the noble gases xenon, argon, neon and helium in an intensity range of \mbox{2--8}$\times$10$^{13}$\,W/cm$^2$ and observed an asymmetry of the photoelectron images along the beam propagation axis with respect to our reference, which is the center spot as marked in Fig.~\ref{fig2}(a). The center spot location corresponds to low-energy electrons that originate from highly excited states that remain after the interaction with the laser pulse~\cite{Smeenk2011,Nubbemeyer2008,Eichmann2013}. The electric field of the spectrometer ranges from 0.5--1\,kV/cm and can thus field-ionize excited states with a binding energy that corresponds to a principal quantum number of $n$=21 or higher~\cite{Gallagher1988}. As these electrons do not gain kinetic energy in the detector plane and do not interact with the laser pulse after ionization, they can be used as a reference point for zero momentum of the photoelectrons~\cite{Smeenk2011}.

\begin{figure*}[t!]
\includegraphics[width=\textwidth]{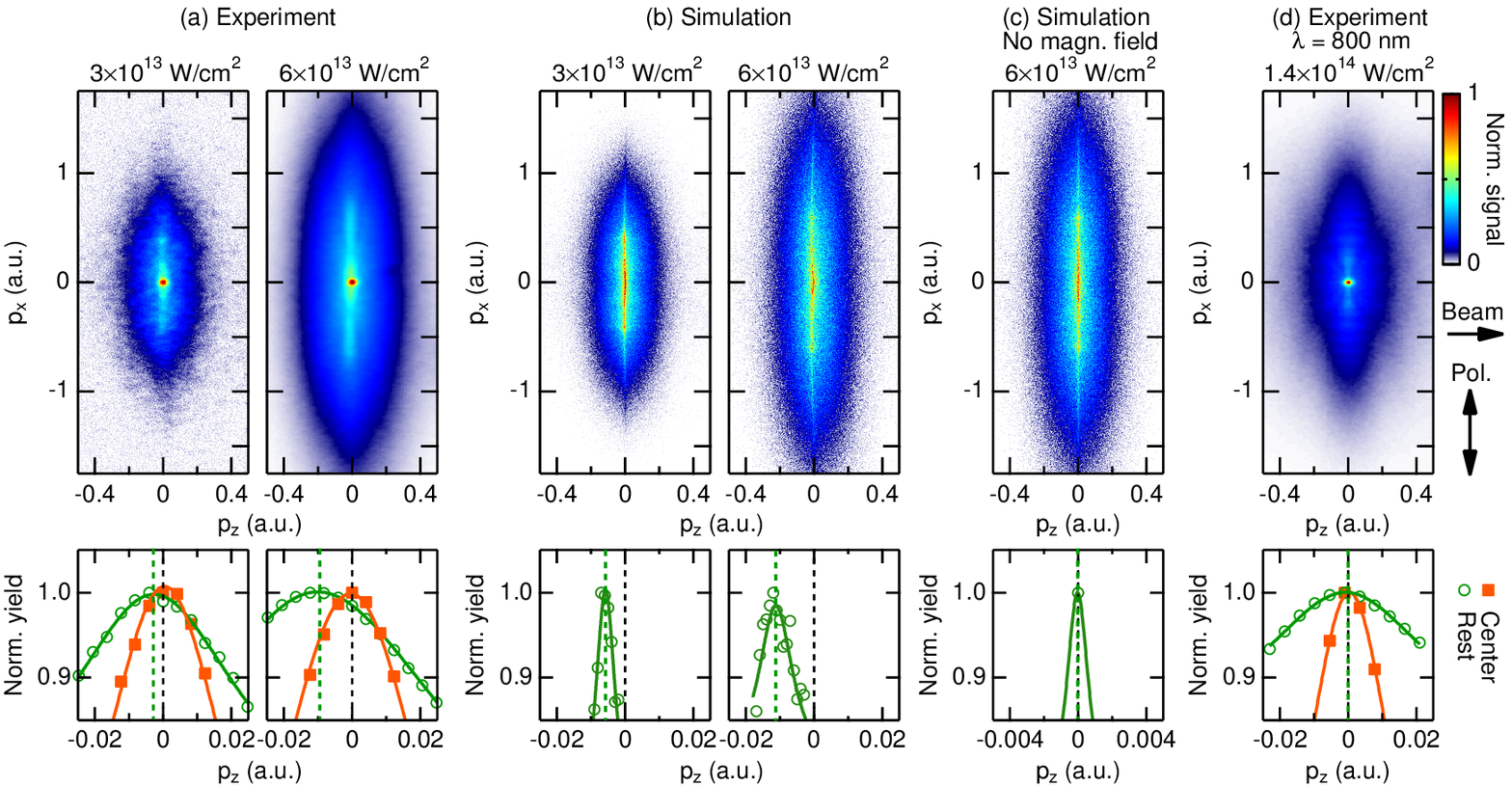}
\caption{\label{fig3}PMDs and their projections recorded in xenon and helium at different intensities. Upper figures: measured and simulated PMDs. Lower figures: projections of these PMDs onto the beam propagation, as used to extract the momentum offset (see Fig. 2). (a) Experimentally-measured PMDs of xenon recorded at 3 and 6$\times$10$^{13}$\,W/cm$^2$. (b) Corresponding simulated PMDs at same intensities than (a) reproducing the negative offset through the combined influence of the Coulomb potential and the full electro-magnetic laser field. (c) Simulated PMD excluding the magnetic field. The projection exhibits no offset. (d) Measured PMD of helium at 800\,nm and an intensity of 1.4$\times$10$^{14}$\,W/cm$^2$. Due to the shorter wavelength, the dipole approximation is valid and no offset is visible.}
\end{figure*}

In order to quantify the asymmetry in the experiment we projected the two-dimensional PMDs onto the axis of beam propagation z and extracted the offset of the peak of the projected distribution with respect to the central reference spot as depicted in Fig.~\ref{fig2}(a). For that reason the PMDs were split along the laser polarization direction x into a central slice with a width of $\Delta p_\text{x}=0.05$\,a.u. (to isolate the central spot), and the two outer regions (that exhibit the offset of the peak in beam propagation direction). As illustrated in Fig.~\ref{fig2}(b) the positions of the maxima were extracted by fitting the peak regions in a range of $\Delta p_\text{z}\approx0.05$\,a.u. with a Lorentzian function in each case. Here the peak of the central slice simply defines $p_\text{z}$=0\,a.u. so we further concentrate our analysis on the offset of the peak of the rest of the PMDs as function of laser intensity. The error of this procedure was estimated from the camera pixel size. The intensity was calibrated via the longitudinal width of the PMDs on the basis of semiclassical calculations described in the following section. In order to prevent the influence of interferences that occur for linear polarization, this longitudinal width was calibrated with measurements and simulations performed for circularly polarized light.

For comparison with the data we performed classical trajectory Monte Carlo (CTMC) simulations of electrons using a semiclassical two-step model~\cite{Gallagher1988a,Corkum1993,Kulander1993,LindenvandenHeuvell1989}. As initial condition, we used the tunnel exit as it is calculated in parabolic coordinates~\cite{Fu2001,Pfeiffer2012} with the ionization rate and the initial momentum distribution from the ADK-theory~\cite{Ammosov1986,Delone1991}.
Despite theoretical models that describe tunnel ionization beyond the dipole approximation~\cite{Klaiber2013,Yakaboylu2013,Milosevic2002}, the validity of this model, that provides our initial conditions for the CTMC simulations, was questioned~\cite{Reiss2008,Reiss2014}. Thus, we tested the robustness of our simulation results against variations of the spatial starting point for the propagation and variations in the ionization rate considerably beyond the change expected for relativistic tunneling~\footnote{We varied the exit point between 60 and 300\% and the ionization rate by at least 5 orders of magnitude.}.
As the outcome was found to be robust against these variations, we can exclude any significant influence due to deviations from the initial conditions. The robustness of our results against details in the geometry or dynamics of the ionization step is further emphasized by the similarity of the data for the different gas species, whose ionization threshold varies considerably. Therefore, we can conclude that the dominant contribution to the observed asymmetry of the PMDs rather originates from the propagation of the liberated electrons under the influence of the combined Coulomb and laser field.

In our model, the magnetic field component of the laser pulse is fully included during the propagation in the combined fields of the laser pulse and the residual ion. Each electron trajectory was propagated until the end of the pulse and the asymptotic momenta were calculated via Kepler's analytical formula~\cite{Shvetsov-Shilovski2012,Shvetsov-Shilovski2009}.
To circumvent numerical problems with the $1/r$ Coulomb potential we filter out electrons that come closer than 0.5~a.u. to the parent ion. This just affects as few as ~0.25\% of the trajectories and was verified not to alter the outcome of the simulation.
For each laser intensity, 10$^6$ trajectories were calculated and subsequently binned in momentum space with a bin size of 10$^{-3}$\,a.u.. In analogy to the procedure with the experimental data, the maxima of the resulting photoelectron images projected onto the beam propagation direction were identified by Lorentzian fits of the central part of the momentum distributions. We would like to mention that the central spot in the simulated PMDs is absent since we did not include the field ionization of highly excited states by the spectrometer field. However in the simulations, the reference for zero momentum is intrinsically known. For the simulations the error was estimated from the bin size used.

In the experiment, the extracted offset of the peak of the PMDs in beam propagation direction shows a clear trend with respect to intensity: in the studied intensity range we observe an increase of the offset for higher intensities. This behavior is also directly visible in the photoelectron momentum images [Fig.~\ref{fig3}(a)]. Furthermore, we observe that the offset is shifted towards negative values on the beam propagation axis, i.e.~opposite to the beam propagation direction. This behavior appears to be counterintuitive as it contradicts the expectation for a free electron, i.e.~without any influence of the Coulomb potential: the behavior of a free electron is expected to be governed by the radiation pressure that is exerted onto it by the Lorentz force. This picture was also utilized in Ref.~\cite{Smeenk2011} for the interpretation of an observed shift of the photoelectron momentum distributions in the beam propagation direction. In contrast, we show that the behavior in the case of linearly polarized light is caused by the influence of the magnetic field and the Coulomb potential. In this case, the electron can be driven back to the ion core by the laser field and can interact with the ion's Coulomb potential~\cite{Corkum1993,Kulander1993}.
As a simplified intuitive picture for the observed asymmetric momentum distribution, one might think of electrons first being pushed in the beam propagation direction by the magnetic field, and then being scattered in the opposite direction by the Coulomb potential when the electrons subsequently pass by the parent ion.

\begin{figure}[t]
\includegraphics{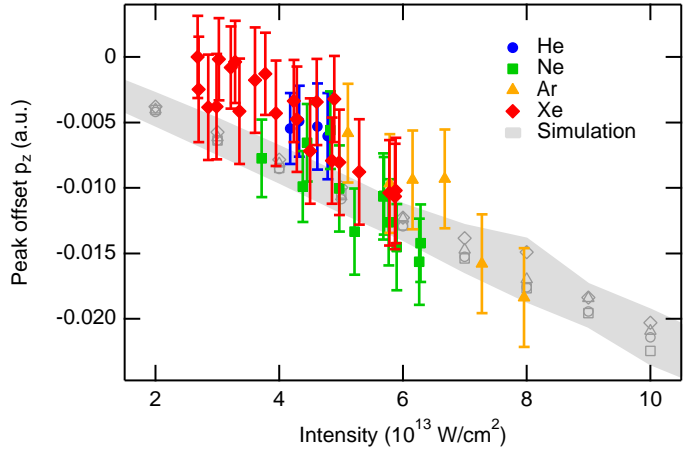}
\caption{\label{fig4}Extracted peak offsets along the laser propagation direction as function of laser intensity for different target gases from experiment (filled markers) and semiclassical simulation (open markers). The points show a clear trend towards increasing negative offsets (i.e.~opposite to the beam propagation direction) for increasing laser intensities. The uncertainties are indicated as error bars and the gray shaded area, respectively.}
\end{figure}

To explain our observations, the experimental data are compared to the CTMC calculations including both the magnetic field component of the laser pulse and the Coulomb potential of the residual ion. We observe a good agreement between the simulated PMDs and the experimental data [Fig.~\ref{fig3}(b)]. When the magnetic field component is neglected in the calculations, the asymmetry along the beam propagation direction vanishes [Fig.~\ref{fig3}(c)]. In order to rule out the possibility that our observed momentum shifts were introduced as an experimental artifact, PMDs were recorded in the same geometry at a wavelength of 800\,nm and are shown in Fig.~\ref{fig3}(d). The intensity used in this photoelectron image was 1.4$\times$10$^{14}$\,W/cm$^2$, i.e.~an intensity that is significantly higher then the ones used for our experiments at mid-IR wavelengths. Nonetheless, this photoelectron image does not show any measurable asymmetry. In Fig.~\ref{fig4} the offsets extracted from the experiments with different gases are plotted together with the ones from the simulations. We observe an excellent agreement of the offsets between experiment and our simple semiclassical theory. For our parameter range, we see an increase of the momentum offset with increasing intensity. Our data demonstrate that the offset is not sensitive to the target gas within the sensitivity limits of the experiment.

A further aspect of strong-field ionization beyond the dipole approximation that has been discussed in recent articles is a possible momentum transfer of the order $I_\text{p}/c$ onto either the ionized electron, the ion or the electron-ion system. In the work of Smeenk \textit{et al.}~\cite{Smeenk2011}, the authors concluded that a momentum $I_\text{p}/c$ is transferred to the electron-ion system before the electron dynamics is governed mainly by the laser field. Other works suggest an initial kick of the electron in the direction of the beam~\cite{Klaiber2013,Yakaboylu2013}. From comparison of simulations where we included a momentum kick of the electron in z-direction with the ones without, we find that in the case of linearly polarized light our measurements are not sensitive enough to resolve consequent signatures in the photoelectron spectra. 

In conclusion, we observed the breakdown of the dipole approximation in its long-wavelength limit for moderate laser intensities in the mid-IR. We showed that for our experimental parameters, the electron dynamics is significantly influenced not only by the magnetic field component of the laser field but also by the Coulomb potential of the parent ion. The action of the Coulomb potential yields rather complicated electron dynamics which challenge the previously-used radiation pressure picture. Thus, concepts~\cite{Klaiber2006,Chirila2002} to compensate for non-dipole effects need to be revisited to take the Coulomb field into account. As the results from our simulations are largely robust against the starting conditions, we conclude that our observations are mainly induced during the quasi-classical dynamics in the continuum which obstructs a direct insight into the nature of the initial ionization step. Our results pose new challenges for the theoretical description of strong-field processes in the long wavelength limit, which is presently of high interest in this research field. However, our findings also open up new possibilities for studying the response of the target system to the magnetic field component inherently present in the laser pulse.

\begin{acknowledgments}
This research was supported by the NCCR MUST, funded by the Swiss National Science Foundation (SNSF) and by the SNSF through grant \#200020\_144365/1. It was supported by a Marie Curie International Incoming Fellowship within the 7th European Community Framework Programme and has been co-financed under FP7 Marie Curie COFUND.
\end{acknowledgments}

\bibliography{DipoleBreakdownRef}

\end{document}